\documentclass[journal]{IEEEtran}
\usepackage{amsmath,amsfonts}
\usepackage{array}
\usepackage{textcomp}
\usepackage{stfloats}
\usepackage{url}
\usepackage{verbatim}
\usepackage{graphicx}
\usepackage{cite}
\newtheorem{theorem}{Theorem}
\newtheorem{myDef}{Definition}

\newtheorem{lemma}{Lemma}
\usepackage{dsfont}
\usepackage{enumerate}
\usepackage{algorithm}
\usepackage{algpseudocode}

\usepackage[caption=false,font=footnotesize,labelfont=rm,textfont=rm]{subfig}

\begin{document}

\title{Efficient Core-selecting Incentive Mechanism for Data Sharing in Federated Learning}

\author{Mengda Ji, Genjiu Xu, Jianjun Ge, Mingqiang Li

\thanks{This work was supported by the National Key Research and Development Program of China under Grant No. 2021YFA1000402 and the National Natural Science Foundation of China under Grant No. 72071159. \emph{(Corresponding author: Genjiu Xu.)}}

\thanks{Mengda Ji is with the Unmanned System Research Institute, Northwestern Polytechnical University, Xi'an 710072, China and the International Joint Research Center on Operations Research, Optimization and Artificial Intelligence, Xi'an 710129, China.}

\thanks{Genjiu Xu is with the School of Mathematics and Statistics, Northwestern Polytechnical University, Xi'an 710129, China and the International Joint Research Center on Operations Research, Optimization and Artificial Intelligence, Xi'an 710129, China. E-mail: xugenjiu@nwpu.edu.cn.}

\thanks{Jianjun Ge and Mingqiang Li are with the Information Science Academy, China Electronics Technology Group Corporation, Beijing 100086, China.}}

%\markboth{IEEE TRANSACTIONS ON COMPUTATIONAL SOCIAL SYSTEMS}%
%{Shell \MakeLowercase{\textit{et al.}}: IEEE TRANSACTIONS ON COMPUTATIONAL SOCIAL SYSTEMS}

%\IEEEpubid{0000--0000/00\$00.00~\copyright~2023 IEEE}

\maketitle
\begin{abstract}
Federated learning is a distributed machine learning system that uses participants' data to train an improved global model. In federated learning, participants collaboratively train a global model, and after the training is completed, each participant receives that global model along with a payment. Rational participants try to maximize their individual utility, and they will not input their high-quality data truthfully unless they are provided with satisfactory payments based on their contributions. Furthermore, federated learning benefits from the cooperation of participants. Accordingly, how to establish an incentive mechanism that both incentivizes inputting data truthfully and promotes stable cooperation has become an important issue to consider. In this paper, we introduce a data sharing game model for federated learning and employ game-theoretic approaches to design a core-selecting incentive mechanism by utilizing a popular concept in cooperative games, the core. In federated learning, the core can be empty, resulting in the core-selecting mechanism becoming infeasible. To address this issue, our core-selecting mechanism employs a relaxation method and simultaneously minimizes the benefits of inputting false data for all participants. Meanwhile, to reduce the computational complexity of the core-selecting mechanism, we propose an efficient core-selecting mechanism based on sampling approximation that only aggregates models on sampled coalitions to approximate the exact result. Extensive experiments verify that the efficient core-selecting mechanism can incentivize inputting high-quality data truthfully and stable cooperation, while it reduces computational overhead compared to the core-selecting mechanism.
\end{abstract}


\section{Introduction}
\IEEEPARstart{I}{n recent years}, the success in the field of machine learning has been mainly attributed to the collection and use of massive amounts of data. Although the demand for data in machine learning is growing, collecting and using data are still difficult. On the one hand, data owners, or participants in federated learning, have recognized the worth of their data, so they are reluctant to provide it for free. On the other hand, privacy protection regulations such as the General Data Protection Regulation (GDPR) impose strict restrictions on access to private data \cite{gdpr}. To address the problem of privacy protection, federated learning has been proposed, which is a cooperative framework based on collaborative training and model sharing \cite{yang2019federated}. In federated learning, the data of participants is stored locally and used to train local models. Then these local models are aggregated by a server to create a global model. Since federated learning does not require participants to upload data to the server, it prevents possible privacy leaks during the uploading, saving, and model training processes. In recent years, federated learning has attracted attention in many fields such as healthcare, finance, and the Internet of Things \cite{rieke2020future, li2021survey, lu2019blockchain}.

In federated learning, economic incentives are crucial for motivating participants to cooperatively train a global model \cite{zhan2021survey}. In practice, participants are reluctant to truthfully input their worthy data unless they receive satisfactory payments for their high-quality data. In addition, the global model relies on the stable cooperation of all participants and the server. Therefore, a desirable incentive mechanism should also take into account cooperative contributions and truthfulness. Along with the development of federated learning, research on incentive mechanisms has attracted many scholars’ interests \cite{tu2022incentive, shi2022fedfaim, lu2022truthful}. Many methods are used to design incentive mechanisms for federated learning, including auction \cite{zeng2020fmore, jiao2020toward, deng2021fair}, contract theory \cite{kang2019incentive}, and matching theory \cite{lim2021towards}. Game theory investigates strategic interactions among players or agents, which provides a framework to analyze the strategic behaviors of participants in federated learning. Therefore, more and more incentive mechanisms have used game-theoretic methods to design incentive mechanisms in federated learning. Such methods include Stackelberg game \cite{sarikaya2019motivating, zhan2020learning}, evolutionary game \cite{luo2022strategic, cheng2021dynamic}, VCG-based (Vickrey-Clarke-Groves) mechanisms \cite{kantarcioglu2010incentive, nix2011incentive, cong2020vcg, cong2020game, zhang2021faithful}, and Shapley value \cite{wang2019measure, sim2020collaborative, ghorbani2019data, wang2020principled}.

%\IEEEpubidadjcol
Many existing studies in federated learning and machine learning focus on designing incentive mechanisms to provide economic incentives for participants. The VCG (Vickrey–Clarke–Groves) mechanism is famous in game theory and mechanism design theory because it satisfies the properties of incentive compatibility and individual rationality. Due to its desirable properties, the VCG mechanism motivates participants to report their private information truthfully. Therefore, the VCG-based incentive mechanism has been widely used in machine learning and federated learning \cite{kantarcioglu2010incentive, nix2011incentive, cong2020vcg, cong2020game, zhang2021faithful}. To reduce computational complexity, a faithful federated learning (FFL) mechanism was proposed to compute the VCG-like payment via an incremental computation \cite{zhang2021faithful}. It effectively reduces computational complexity, making it suitable for large-scale training scenarios. The utility of participants under the VCG-based mechanism is equal to their marginal contribution, which motivates them to input high-quality data. In addition to computing the VCG payment, allocating payments or profits based on cooperative contributions to promote stable cooperation is also a challenge in federated learning. % The VCG-based mechanism does not consider some desirable properties from the perspective of cooperative game theory, which results in some participants being dissatisfied with the profits and even dropping out.

Cooperative game theory offers various solution concepts to determine how benefits should be allocated. The Shapley value is a classic solution in cooperative game theory with desirable properties and interpretability, which has been widely adopted in incentive mechanisms for machine learning and federated learning \cite{wang2019measure, song2019profit, jia2019towards, ghorbani2019data, wang2020principled, sim2020collaborative}. These methods allocate payments according to the marginal surplus of each participant, which reflects the combined worth of participants' data. Except for Shapley value, the core is also a widely used solution concept in cooperative game theory \cite{gillies1953some, scarf1967core, telser1994usefulness}. It represents a stable and fair allocation set that satisfies the properties of individual rationality and coalitional rationality, which is different from Shapley value. The allocation element in the core ensures that no player or coalition has the motivation to deviate from cooperation. Due to its desirable properties, the core has been widely applied in many fields such as spectrum auctions, crowdsourcing, and transportation scenarios \cite{zhu2014core, hu2018dtcs, james2018core}. Recently, the core has also been used in federated learning to evaluate the cooperative contributions of participants \cite{ray2022fairness, donahue2021optimality}. These methods fully analyze the existence of the core and provide a fair allocation scheme. Nevertheless, the core is a set solution that cannot provide a unique optimal solution, and the advantages of the VCG payment or marginal contribution have not been fully considered by these methods. Core-selecting mechanisms, often referred to as core-selecting auctions, are designed to ensure that the resulting allocation is in the core of the associated cooperative game \cite{day2012quadratic}. Therefore, we will use the core-selecting mechanism and adopt some improvements to address potential issues that may arise in federated learning. Additionally, it is worth noting that determining the exact core in federated learning also faces computational difficulties because it requires computing additional $2^n$ models \cite{yan2021if}. 
% In addition, the above studies allocate the profits of global model to all participants, but their utility functions are not fully applicable in some scenarios of federated learning. Generally, the parameters of the global model are updated jointly by all participants, and in some scenarios, participants will share the same global model in the update process. Therefore, receiving the global model is also a non-monetary gain for each participant if the global model performs better than the local models. 

As shown in Table \ref{table}, we give the comparison table to highlight the differences of existing incentive mechanisms. 

\begin{table}[ht]\label{table}
	\centering
	\setlength{\tabcolsep}{2.5pt}
	\caption{The differences of existing incentive mechanisms.}
	\begin{tabular}{lllll}
		\hline
		& \begin{tabular}[c]{@{}l@{}}Cooperative\\ incentive\end{tabular} & \begin{tabular}[c]{@{}l@{}}Truthful\\ incentive\end{tabular} & \begin{tabular}[c]{@{}l@{}}Additional\\ aggregation\end{tabular} & \begin{tabular}[c]{@{}l@{}}Additional\\ training\end{tabular} \\ \hline
		\cite{jia2019towards, ghorbani2019data, wang2020principled} & Shapley value          & \textbackslash{} & \textbackslash{}       & Yes                 \\
		\cite{ray2022fairness, donahue2021optimality}               & core                   & \textbackslash{} & Yes                    & \textbackslash{}    \\
		\cite{yan2021if}                                            & least core             & \textbackslash{} & \textbackslash{}       & Yes                 \\
		\cite{cong2020vcg}     & \textbackslash{} & VCG-based & Yes & \textbackslash{} \\
		\cite{kantarcioglu2010incentive, nix2011incentive}          & \textbackslash{}       & VCG-like & \textbackslash{}       & Yes                 \\
		\cite{zhang2021faithful}                                    & \textbackslash{}       & VCG-like & Yes                    & \textbackslash{}    \\
		This paper             & strong $\epsilon$-core & VCG-like & Yes                    & \textbackslash{}    \\ \hline
	\end{tabular}
\end{table}

In this paper, we propose an efficient core-selecting incentive mechanism for federated learning. First, we introduce a data sharing game for federated learning. Then we introduce a core-selecting incentive mechanism that combines the advantages of both the VCG-like payment and the core, which can promote stable cooperation among players and also motivate participants to input their high-quality data. Different from the classical core-selecting mechanism, we adopt a relaxation on the core to deal with the randomness in federated learning. Since the core-selecting incentive mechanism requires exponential time to aggregate and evaluate additional models in federated learning, it is difficult to compute the exact solution. To address this, we propose an efficient core-selecting mechanism based on sampling approximation that significantly reduces computational complexity. This mechanism adjusts the aggregation weight of participants based on their historical contributions to avoid the impact of low-quality data. In addition, we verify the theoretical results of the proposed mechanism in experiments. Further experiments show that the efficient core-selecting mechanism can motivate participants to truthfully input high-quality data and promote stable cooperation, while it reduces computational overhead compared to the core-selecting mechanism.

% In conclusion, the proposed efficient core-selecting mechanism balances the properties of cooperative stability, incentive compatibility, and computational feasibility, making it applicable in practical scenarios. 

Our contributions are as follows:
\begin{enumerate}
	\item We introduce a data-sharing game for federated learning to study participants' strategic behaviors. Based on the outcomes of this data-sharing game, we define the characteristic function and the core to measure participants' cooperative contributions.
	
	\item We employ a core-selecting mechanism for federated learning, which aims to find the optimal payment based on the core to incentivize participants' cooperation. Due to the impact of low-quality data or overfitting on the global model, the core may be empty. To address this issue, our core-selecting mechanism employs a relaxed version of the core, strong $\epsilon$-core, and minimizes the benefits of inputting false data. 
	
	\item To avoid the core-selecting incentive mechanism implementing in an exponential time to aggregate and evaluate additional models in federated learning, we propose an efficient core-selecting incentive mechanism based on sampling approximation that significantly reduces computational complexity. This incentive mechanism aggregates new global models based on the historical contributions of participants.
\end{enumerate}

The rest of this paper is organized as follows. Section \ref{Background and Problem Formulation} introduces the background of federated learning and formulates the problem. Section \ref{Truthful incentive mechanism for Federated Learning} introduces the truthful incentive mechanism for federated learning. The core-selecting mechanism is proposed in Section \ref{Core-selecting Incentive Mechanism for Federated Learning}. The efficient core-selecting mechanism and its theoretical results are proposed in Section \ref{Efficient Core-selecting Incentive Mechanism based on Sampling Approximation and Reputation}. Experiments are given in Section \ref{Experiments} to illustrate the performance of our method. Section \ref{Conclusion} concludes this paper.

\section{Background and Problem Formulation}\label{Background and Problem Formulation}

We first introduce the background of federated learning in Section \ref{Federated Learning}, and then introduce the necessary notions and definitions in Section \ref{Problem Formulation}. 

\subsection{Background of Federated Learning}\label{Federated Learning}

\begin{figure}[htbp]
\centering
\includegraphics[width=3in]{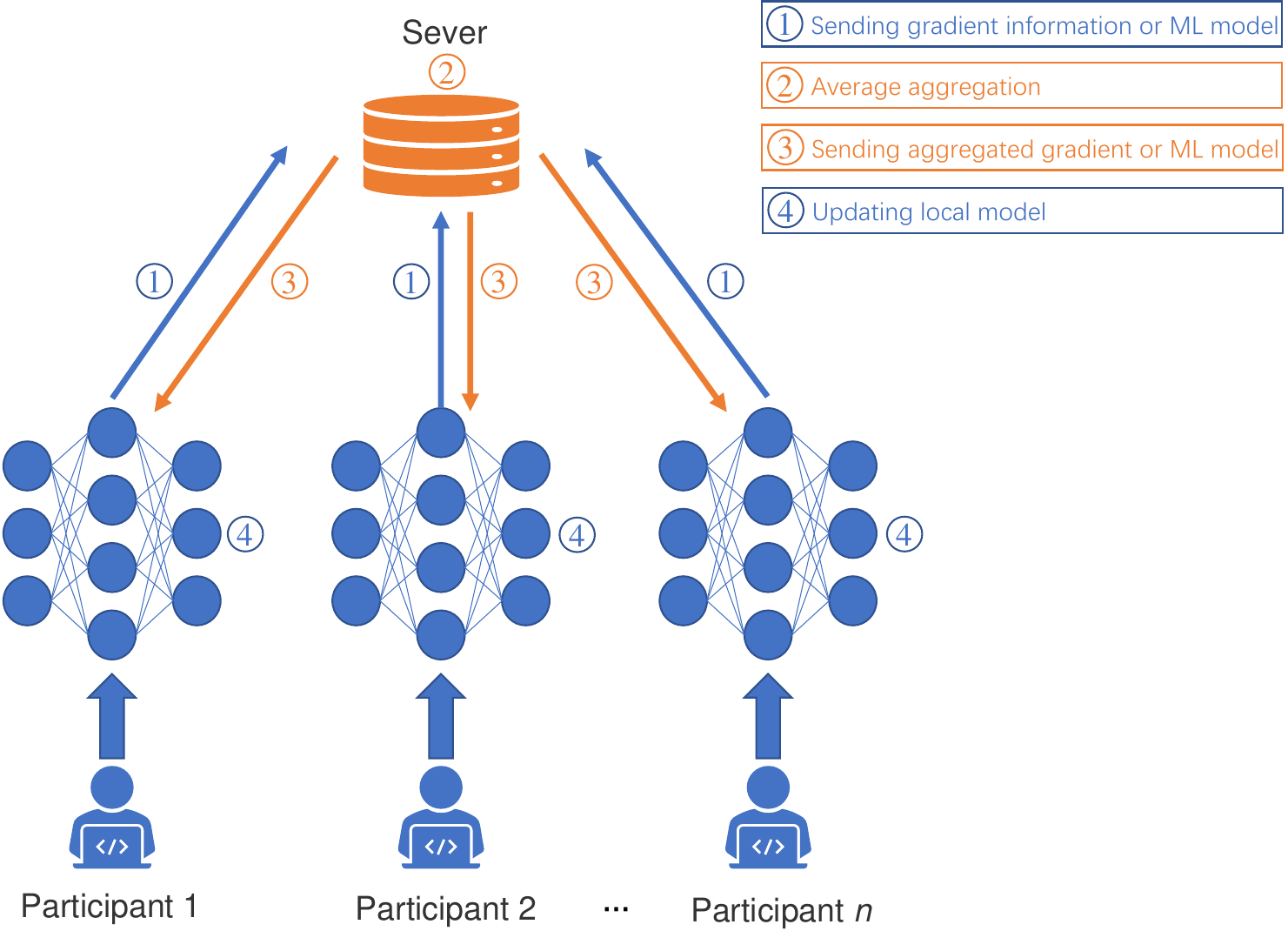}
\caption{The system of federated learning.}
\label{FL_architecture_diag}
\end{figure}

We introduce a typical federated learning system, as shown in Fig. \ref{FL_architecture_diag}. In this system, $n$ participants with data collaboratively train a machine learning model with the help of a server \cite{aono2017privacy}. The training process of federated learning system typically consists of the following four steps.

\begin{enumerate}[Step 1:]
 
\item Participants compute the model gradients locally. Mask gradients information with homomorphic encryption \cite{aono2017privacy}, differential privacy \cite{abadi2016deep}, or secret sharing techniques \cite{bonawitz2017practical}, and send the masked result to the server.

\item The server performs a secure aggregation operation on gradients information from participants, such as using weighted averaging based on data size \cite{yang2019federated, bonawitz2017practical}.

\item The server sends back the aggregated results (gradients information) to participants.

\item Participants update their local model parameters with the gradient results from the server.
\end{enumerate}

The process of above steps will continue until the loss function converges. This system is independent of specific machine learning algorithms (logical regression, deep neural networks, etc.), and all participants will share the final model parameters. 

In the above steps, the participant sends gradient information, and the server aggregates the gradient information using a weighted average method. Therefore, this method is called gradient averaging \cite{mcmahan2017communication, yu2019distributed}. In addition to sharing gradient information, participants in federated learning can also share models. Participants calculate model parameters locally and send them to the server \cite{phuong2019privacy}. The server aggregates the received model parameters (for example, calculates a weighted average) and then sends the aggregated model parameters to the participants. This method is called model averaging, as shown in Algorithm \ref{FedAvg}. Experiments have shown that model averaging is equivalent to gradient averaging, so both are called federated averaging \cite{mcmahan2017communication}. 

\begin{algorithm}[htbp]
	\caption{FedAvg}
	\begin{algorithmic}	
		\State Initialize global model $\theta^0$.
		\For{each round $t = 1,2,\cdots,T$}
		\For{each participant $i \in N$ \textbf{in parallel}}
		\State $\theta_{i}^{t} = \text{ParticipantUpdate}(i, \theta^{t-1})$
		\EndFor
		\State Update global model $\theta^{t} \leftarrow \sum_{i \in N} \frac{1}{|N|} \theta_{i}^{t}$
		\EndFor
		
		\State \textbf{$\text{ParticipantUpdate}(i, w)$} \% \emph{Run on participant $i$}
		\State $Z \leftarrow$ (split $D_i$ into batches of size $B$)
		\For{each local epoch $e = 1, \cdots, E$}
		\For{batch $z \in Z$}
		\State $\theta \leftarrow \theta - \eta \nabla \mathcal{L}(\theta; z)$
		\EndFor
		\EndFor
		\State return $\theta$ to server
	\end{algorithmic}
	\label{FedAvg}
\end{algorithm}

It is important to note that federated learning requires stable cooperation among participants and the server. In addition, on the above system, participants are not supervised when training local models, so no one else knows whether participants are using truthful data or generated false data. In order to promote stable cooperation and encourage training with truthful data, it is necessary to design an incentive mechanism for federated learning to determine reasonable payments for participants.

\begin{figure}[!h]
	\centering
	\includegraphics[width=3in]{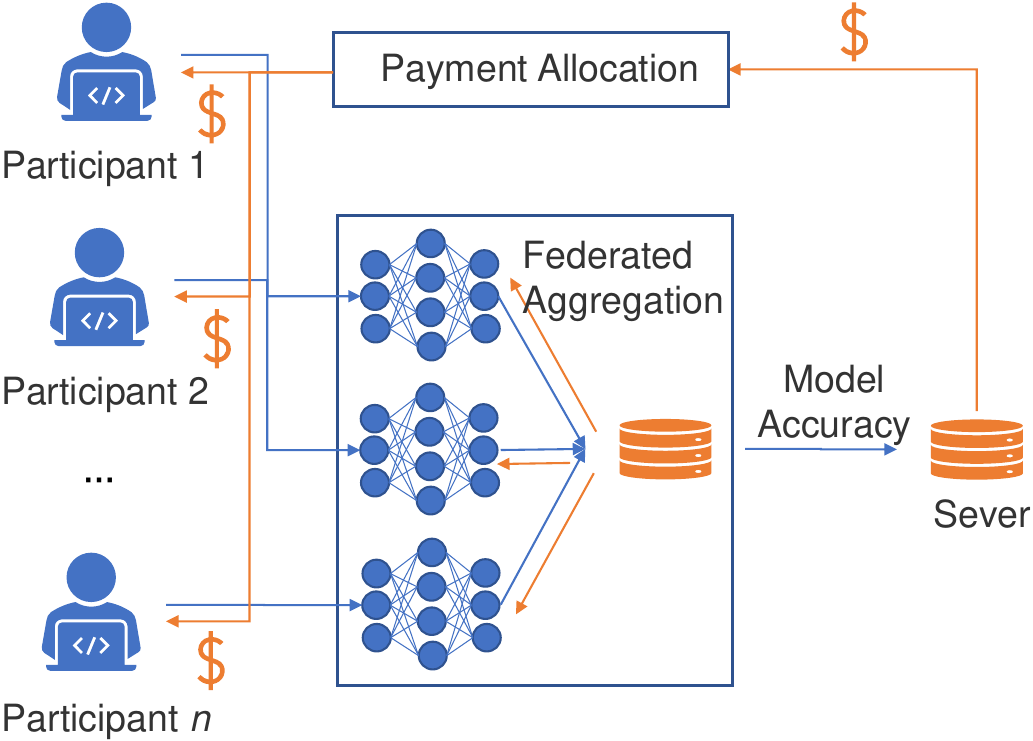}
	\caption{The framework of incentive mechanism for federated learning.}
	\label{The framework of incentive mechanism}
\end{figure}

In this paper, we consider an incentive mechanism framework for federated learning, as shown in Fig. \ref{The framework of incentive mechanism}. The operator sets up a server responsible for coordinating participants to implement a federated learning algorithm, such as the FedAvg algorithm. The server measure each participant's contribution and provide them with payments when the global model is given at each round. In the following sections, we will study some desirable incentive mechanisms based on this framework.

\subsection{Problem Formulation}\label{Problem Formulation}
In order to analyze federated learning from the perspective of game theory, a data sharing game environment outlining federated learning algorithm is proposed in this section. We denote the set of all participants as $N = \{1,\cdots, n\}$, and each participant's dataset is denoted as $D_i = \{x_{ij}, y_{ij}\}_{j=1}^m$. In each round $t$, each participant $i$ updates their local model and the global model using their inputted dataset $D_i$. We will employ game theory to analyze the strategic behavior of these $n$ participants in each round.

\begin{enumerate}
	\item Players: Participants $N = \{1,\cdots,n\}$ and the server $0$.
	
	\item Type: Each participant $i \in N$ has a dataset $D_i$ to input during the local training process, which can be called player $i$'s type in game theory. Note that $D_i$ is $i$'s private information, not known to others. 
	
	\item Strategy: Each player $i \in N$ selects $\hat{D}_i$ as its input, which can be truthful, i.e., $\hat{D}_i = D_{i}$, or untruthful, i.e., $\hat{D}_i \neq D_{i}$. In addition, each participant can also choose to quit ($\hat{D}_i = \emptyset$). We denote $D = \cup_{i\in N}D_i$ as the original dataset, and $\hat{D} = \cup_{i\in N}\hat{D}_{i}$ as the selected dataset. 
	
	\item Each player $i\in N$ inputs its selected dataset to update its local model $F(\hat{D}_i)$, and then a global model is aggregated from these $n$ local models, which is denoted as $F(\hat{D})$. We can write the accuracy function $A(F(\cdot))$, where $F(\cdot)$ represents a federated learning algorithm that maps an inputted dataset onto a global model and $A(\cdot)$ is some accuracy metric applied to the model.
	
	\item As the global model is shared among players during the update process, each participant will receive the parameters of the global model after this round. Each player $i \in N$ will gain potential benefits from the global model if it is better than its local model $F(D_i)$. To characterize the immediate benefits for the current round, we define the valuation function of player $i$ as
	\begin{equation}\label{valuation function}
		v_{i}(F(\hat{D}), D_i) = E[\max \{h_i(F(\hat{D})) - h_i(F(D_{i})), 0\}],
	\end{equation}
	where $h_i$ is a pricing function for a given model, defined as a monotonically increasing function of the model accuracy. The specific form of $h_i$ is given by the participants and operator. For simplicity, we set $h_i(\cdot) = k_i A(\cdot)$, where $k_i > 0$ is a constant number that represents $i$'s preference for a model. $v_i(F(\hat{D}), D_i)$ represents player $i$'s benefits valuation for the global model $F(\hat{D})$ when player $i$'s original dataset is $D_i$. 
	
	\item Each player $i \in N$ will receive a payment $p_i$, which is to be determined. Then player $i$'s utility function is defined as
	\begin{equation}
		u_{i}(F(\hat{D}), D_i) = v_{i}(F(\hat{D}), D_i) + p_{i}.
	\end{equation}
	
	\item The server has a budget $b_0$ to pay for all participants. Thus, the utility of player $0$ is defined as $u_0 = b_0 - \sum_{i \in N} p_i$.
\end{enumerate}

Note that the initial update of a participant's local model is based on its own dataset, while subsequent updates are based on the global model aggregated in the previous round. The valuation function (\ref{valuation function}) is used to evaluate the potential benefits in each round. For participant $i$, the cumulative valuation of all rounds represents the improvement of the global model compared to $i$'s initial model.

A cooperative game is determined by a player set $M$ and a characteristic function $w$. The characteristic function $w$ assigns a numerical value $w(S)$ to each subset $S \subseteq M$ (also called a coalition) of the player set, representing the total worth that the coalition can obtain.

In order to study the cooperative behavior of players using cooperative game methods, we need to define a characteristic function based on the outcome of the data sharing game. 

In federated learning, players gain benefits from the global model. Hence, the worth generated by all players $N \cup \{0\}$, can be written as
\begin{equation}
	w(N) = \sum_{i \in N} u_{i}(F(\hat{D}), \hat{D}_i) + u_0,
\end{equation}
where $w(N) = \sum_{i \in N}v_{i}(F(\hat{D}), \hat{D}_i) + b_0$ also holds.

Assuming the model training is restricted to coalition $S \subseteq N$, and coalition $S$ gets a global model $F(\hat{D}_S)$ where $\hat{D}_S = \{\hat{D}_{i}\}_{i \in S}$, then the worth generated by coalition $S \cup \{0\}$ is
\begin{equation}
	w(S) = \sum_{i \in S}v_{i}(F(\hat{D}_S), \hat{D}_i) + b_0.
\end{equation}

Note that we use the notation $w(S)$ instead of $w(S \cup \{0\})$, as the cooperation of federated learning definitely requires the server $0$ and notation $w(S)$ is more convenient to use. Thus, the characteristic function is given as follows.
\begin{equation}\label{characteristic function}
	w(S) = \begin{cases}
		\sum_{i \in S} v_i(F(\hat{D}_S), \hat{D}_i)+b_0, &S \in 2^N \backslash{\emptyset}, \\
		0, &S = \emptyset.
	\end{cases}
\end{equation}
where $2^N$ represents all subsets of $N$. To compute all characteristic values, the server has to aggregate additional models for all coalitions $S \in 2^N \backslash \emptyset$. By inputting these characteristic values, solutions in cooperative games, such as the core, can determine a payoff allocation scheme, thus incentivizing players' cooperative behaviors.

The problem studied in this paper is to design a desirable incentive mechanism based on these $2^n$ characteristic values. Based on our previous discussion, this mechanism is designed to incentivize each player to truthfully input its high-quality dataset, with no player willing to deviate from cooperation. In order to design such a mechanism, the following properties will be considered. 

\begin{itemize}
	\item \textbf{Incentive Compatibility:} All players input their dataset truthfully, which is a Nash strategic equilibrium. In other words, any participant has no motivation to deviate from inputting the original dataset because the utility of inputting original dataset is not less than the utility of inputting false dataset:
	\begin{equation}
		u_{i}(F(D), D_i) \geq u_{i}(F(\hat{D}_i \cup D_{-i}), D_i), \ \ i \in N.
	\end{equation}
	
	\item \textbf{Individual Rationality:} Any player can get non-negative utility when inputting dataset truthfully:
	\begin{equation}
		u_{i}(F(D), D_i) \geq 0, \ \ i \in N.
	\end{equation}
	
	\item \textbf{Coalitional Rationality:} Any coalition has no motivation to split off:
	\begin{equation}
		\sum_{i \in S} u_i(F(D), D_i) + u_0 \geq w(S).
	\end{equation}
	
	\item \textbf{Computational Efficiency:} The incentive mechanism can be implemented in polynomial time.
\end{itemize}

The first two properties described above are intended to encourage players to input dataset truthfully. The third property is to ensure the stability of cooperation among all players and the server.

\section{Truthful Incentive Mechanism for Federated Learning}\label{Truthful incentive mechanism for Federated Learning}
The VCG (Vickrey–Clarke–Groves) mechanism is famous in game theory and mechanism design theory, which satisfies the property of incentive compatibility, i.e., telling the truth is a dominant strategy for all players \cite{narahari2014game}. The VCG mechanism was previously employed by Nix and Kantarcioglu in the context of distributed machine learning \cite{nix2011incentive}, and they have proposed a critical assumption in their work. Meng Zhang et al. have also investigated VCG-based mechanisms in the context of federated learning and proposed a VCG-like payment to approximate VCG payments via an incremental computation. In this section, with minor modifications, we also implement this VCG-based mechanism for our incentive mechanism framework. This truthful incentive mechanism will serve as the foundation for the core-selecting mechanism in the next section.

In federated learning, a commonly used assumption is that each dataset $D_i$ and the test dataset $D_t$ are sampled from a real distribution $\mathcal{D}$. During the training process, participant $i$ actually inputs an untruthful dataset $\hat{D}_i = \{\hat{x}_{ij}, \hat{y}_{ij}\}_{j=1}^{m}$, where $\hat{D}_i$ deviates from the real probability distribution $\mathcal{D}$ and follows a probability distribution with artificial errors, i.e.,  $\mathcal{D} + \delta$.

Let $f = F(D)$ and $\hat{f} = F(\hat{D}_i \cup D_{-i})$, the expected risks of these two models on distribution $\mathcal{D}$ are denoted as $E_{\mathcal{D}}(f) = \int \mathcal{L}(f(x), y)d\mathcal{D}(x, y)$ and $E_{\mathcal{D}}(\hat{f}) = \int \mathcal{L}(\hat{f}(x), y) d\mathcal{D}(x, y)$ where $\mathcal{L}$ is a per-sample loss function. Generally, $E_{\mathcal{D}}(f) \leq E_{\mathcal{D}}(\hat{f})$ holds because model $f$ fits the data points under the real distribution $\mathcal{D}$, whereas model $\hat{f}$ fits the noisy data points. As a result, on the test dataset $D_t \sim \mathcal{D}$, the performance of $F(D)$ will be better than that of $F(\hat{D}_{i} \cup D_{-i})$. Mathematically, this can be expressed as
\begin{equation}\label{assumption}
	E[A(F(D))] \geq E[A(F(\hat{D}_i \cup D_{-i}))] + I(dist(X_i, \hat{X}_i)),
\end{equation}
where $I$ is a non-negative, increasing function and $dist$ is a distance function of the dataset.

This assumption is similar to the assumption proposed by Nix and Kantarcioglu \cite{nix2011incentive}. It means that deviating from a real data will make a bad model's performance more likely. Obviously, a truthful input will not decrease the model accuracy for each player $i$, and more input of truthful data will improve model performance. Nix and Kantarcioglu also mentioned the possible effects of the overfitting phenomenon. In order to verify the validity of this assumption in the context of federated learning, we will test it by inputting a noisy dataset in Section \ref{Experiments}.

The truthful incentive mechanism is easy to operate. Given an input $\hat{D}$, each player $i \in N$ receives a global model $F(\hat{D})$ and a VCG-like payment, which is defined as
\begin{equation}\label{VCG payment}
	p_{i}^{VCG} = \sum_{j \neq i}v_{j}(F(\hat{D}), \hat{D}_j) - \sum_{j \neq i}v_{j}(F(\hat{D}_{-i}), \hat{D}_{j}), \ \ i \in N.
\end{equation}

To compute the payments for all participants, additional $n$ models need to be aggregated and evaluated. Note that it is not strictly a classic VCG mechanism. The difference between this mechanism and the classical VCG mechanism is that the selection of players is not executed here. The reasons are twofold. Firstly, in order to find the optimal model and achieve a selection of players, $2^n$ times of aggregation and $2^n$ times of evaluation for all potential selection schemes are required, which are computationally expensive. Secondly, only after a participant is selected can we obtain its local model and evaluate it. Thus, the selection of participants is not considered in this paper. Although the selection is not performed by this mechanism, it still satisfies incentive compatibility.

\begin{theorem}[Nix and Kantarcioglu \cite{nix2011incentive}]\label{thm1}
	The truthful mechanism satisfies the properties of incentive compatibility and individual rationality.
\end{theorem}
\begin{IEEEproof}[Proof (Incentive Compatibility)]
	Given any input profile $D_{-i}$ of other players, if $i$ inputs $\hat{D}_i$, player $i$'s utility is
	\begin{equation}
		\begin{aligned}
			u_{i}(F(\hat{D}_i \cup D_{-i}),D_i) =& v_{i}(F(\hat{D}_i \cup D_{-i}),D_i) + p_{i}^{VCG}\\
			=& \sum_{j \in N}v_{j}(F(\hat{D}_i \cup D_{-i}),D_j)  \\
			&- \sum_{j \neq i}v_{j}(F(D_{-i}),D_{j}).
		\end{aligned}
	\end{equation}
	
	Similarly, if $i$ inputs truthfully, player $i$'s utility is
	\begin{equation}
		\begin{aligned}
			u_{i}(F(D), D_{i}) =& \sum_{j \in N}v_{j}(F(D),D_{j}) - \sum_{j \neq i}v_{j}(F(D_{-i}),D_{j}). \\
		\end{aligned}
	\end{equation}
	
	In order for incentive compatibility to exist, this requires that
	\begin{equation}
		\sum_{j \in N}v_{j}(F(D),D_{j}) \geq \sum_{j \in N}v_{j}(F(\hat{D}_i \cup D_{-i}),D_j).
	\end{equation}
	
	We assert that $E[\max \{h_i(F(D)) - h_i(F(D_{i})), 0\}] \geq E[\max \{h_i(F(\hat{D}_{i} \cup D_{-i})) - h_i(F(D_{i})), 0\}]$ always holds, since either the last expression is zero, in which case the first expression is greater than or equal to zero, and the last expression is greater than zero, in which case the first expression is greater than or equal to the last expression due to inequality (\ref{assumption}). Therefore, we have $v_{j}(F(D),D_j) \geq v_j(F(\hat{D}_i \cup D_{-i}), D_j)$ and the above inequality holds.
\end{IEEEproof}
\begin{IEEEproof}[Proof (Individual Rationality)]
	To show that the mechanism is individually rational, we only need to show that the mechanism has a utility of at least zero. The utility of player $i$ is
	\begin{equation}
		\begin{aligned}
			u_i(F(D), D_i) =& \sum_{j \in N} v_{j}(F(D),D_j)  -\sum_{j \neq i}v_{j}(F(D_{-i}), D_{j}) \\
			\geq & \sum_{j \neq i} (v_{j}(F(D),D_j)  - v_{j}(F(D_{-i}), D_{j})).
		\end{aligned}
	\end{equation}
	
	Let $\hat{D}_{i} = \emptyset$, according to inequality (\ref{assumption}), we have
	\begin{equation}
		v_{j}(F(D),D_j) \geq v_{j}(F(D_{-i}), D_{j}).
	\end{equation}
	
	Hence, 
	\begin{equation}
		u_i(F(D), D_i) = \sum_{j \neq i} (v_{j}(F(D),D_j)  - v_{j}(F(D_{-i}), D_{j})) \geq 0.
	\end{equation}
\end{IEEEproof}

The VCG-like payment measures the difference between the global model, i.e., $F(\hat{D})$, and the model without his input, i.e., $F(D_{-i})$. In economics, allocating benefits by marginal contributions encourages participants to adopt decisions and behaviors that have a positive impact on team benefits. If participant $i$ inputs data truthfully, its utility will be equal to its marginal contribution $w(N) - w(N\backslash i)$, which encourages participants to input high-quality data to improve $w(N)$. This truthful incentive mechanism, or the VCG-like payment, meets the requirements of the first two properties mentioned in Section \ref{Background and Problem Formulation}.

\section{Core-selecting Incentive Mechanism for Federated Learning}\label{Core-selecting Incentive Mechanism for Federated Learning}
While the truthful incentive mechanism can motivate players to input their high-quality data, it fails to consider cooperation among all players and the server. The core is a classic set solution in cooperative games. In this section, we define the core based on the outcomes of the proposed data sharing game. The core cannot be guaranteed to be nonempty in federated learning. To address this issue, we adopt a relaxed version of the core, strong $\epsilon$-core. Then, we propose a core-selecting mechanism that selects a solution minimizing participants' benefits of untruthful input within the strong $\epsilon$-core.

Let $\pi_{i} = v_i(F(\hat{D}),\hat{D}_i) + p_i$ be the observable surplus of player $i$. It does not involve the private information of players and can be computed by the server. Instead of the true utility $u_i$, the core is defined over surplus $\pi_i$ because the server does not have any knowledge on the players' original data $D_i$, without a guarantee on the incentive compatibility \cite{day2012quadratic}. The true utility and the incentive for participants to input data truthfully will be discussed later. The server’s surplus is defined as $\pi_0 = u_0= b_0 - \sum_{i \in N} p_i$.

\begin{myDef}[Core]\label{Core}
	A surplus vector $\pi = (\pi_0, \pi_{1}, \cdots, \pi_{n})$ is feasible if $\sum_{i \in N}\pi_i + \pi_0 = w(N)$, which means that all players completely allocate the worth generated in federated learning. A surplus vector is blocked by coalition $S \subseteq N$ if there is another $\pi'$ such that $\pi_k' \geq \pi_k$ for all $k \in S$, and $\pi_0' > \pi_0$. The core is the set of non-negative surplus vectors that are feasible and not blocked by any coalition, which can be mathematically defined as
	\begin{equation}
		\begin{aligned}
			\text{Core}(N) = \{(\pi_0, \pi_1, \cdots, \pi_n) \in \mathbb{R}^{n+1}_{+}|\sum_{i \in N} \pi_{i} + \pi_{0} = w(N), \\
			\sum_{i \in S} \pi_{i} + \pi_{0} \geq w(S), S \subseteq N \}.
		\end{aligned}
	\end{equation}
\end{myDef}

This implies that when the server selects a vector from the core, there is no motivation for any player $k$ to form a coalition $S$ with other players to deviate from coalition $N$ for any possible improvement of their total utility. 

\begin{figure}[htbp]
	\centering
	\includegraphics[width=2.5in]{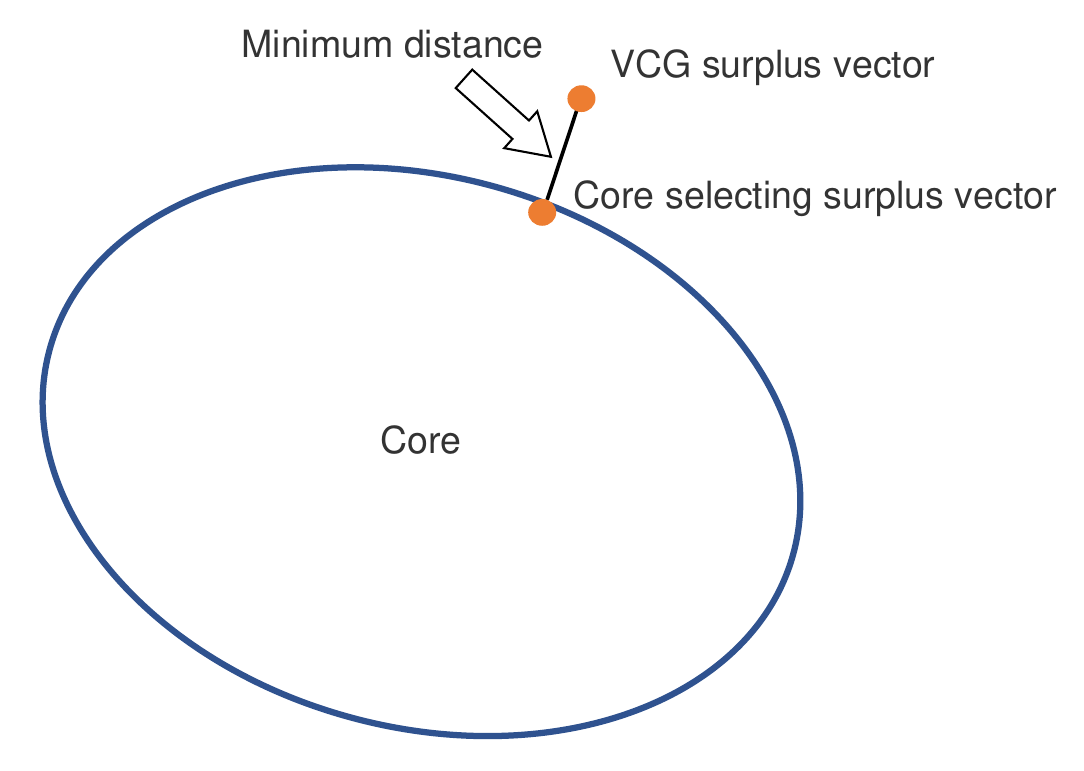}
	\caption{Selecting the optimal surplus vector from the core.}
	\label{Selecting the optimal surplus vector from the core}
\end{figure}

The classical core-selecting mechanism, as shown in quadratic programming (\ref{classical core-selecting}) and Fig. \ref{Selecting the optimal surplus vector from the core}, aims to find a surplus from the core, which is closest to the surplus vector determined by the VCG-like payment, $\pi^{VCG} = v_{i}(F(\hat{D}), \hat{D}_{i}) + p_i^{VCG}$. It combines the advantages of the VCG-based mechanism and the core. 

\begin{equation}\label{classical core-selecting}
	\begin{aligned}
		\min & \ \ \sigma^2 = \sum_{i \in N} (\pi_i - \pi_i^{VCG})^2 \\
		\text{s.t.}
		& \ \ \sum_{i \in N} \pi_i + \pi_0 = w(N), \\
		& \ \ \sum_{i \in S} \pi_i + \pi_0 \geq w(S), \ \ \ \ &S \subseteq N,\\
		& \ \ \pi_i \geq 0, &i \in N \cup \{0\},\\
	\end{aligned}
\end{equation}

\begin{theorem}\label{thm3}
	The core is nonempty if $F(D)$ performs better than $F(D_S)$ for any coalition $S \subseteq N$.
\end{theorem}
\begin{IEEEproof}
	We show that the surplus vector determined by the first price rule is in the core. Given any type profile $D$, for each player $i$, we let $p_i^1 = - g_i(F(D), D_i)$ and then we have $\pi_i^1 = 0$ for all $i \in N$. For any coalition $S \subseteq N$, we have
	\begin{equation}
		\begin{aligned}
			\sum_{i \in N}\pi_i^1 + \pi_0^1 &= \pi_0^1 = b_0 - \sum_{i \in N} p_i^1 \\
			&= \sum_{i \in N} v_i(F(D), D_i) + b_0 \\
			&= w(N),
		\end{aligned}
	\end{equation}
	and
	\begin{equation}
		\begin{aligned}
			\sum_{i \in S}\pi_i^1 + \pi_0^1 &= b_0 + \sum_{i \in N} v_i(F(D), D_i) \\
			&\geq \sum_{i \in S} v_i(F(D_S), D_i) + b_0 \\
			&= w(S). 
		\end{aligned}
	\end{equation}
	
	Therefore, the surplus vector determined by the first price rule is in the core, so the core is nonempty.
\end{IEEEproof}

Intuitively speaking, with more input, $F(D)$ should perform better than $F(D_S)$ and the core should be a nonempty set. However, there are some issues in practice. Firstly, in practical implementation, we cannot repeat training infinitely to accurately approximate the expected value of accuracy. Additionally, model overfitting may also result in $E[A(F(D))] < E[A(F(D_T))]$ and even $w(T) > w(N)$ holding for a coalition $T \subset N$. Both two issues result in an empty core, and the core-selecting cannot be used directly. To address this, we employ a non-empty set solution similar to the core, strong $\epsilon$-core, which adopts a relaxation $\epsilon$ to the core \cite{driessen2013cooperative}. The strong $\epsilon$-core is defined as
\begin{equation}
	\begin{aligned}
		\text{Core}_{\epsilon}(N) = \{(\pi_0, \pi_1, \cdots, \pi_n) \in \mathbb{R}^{n+1}_{+}|\sum_{i \in N} \pi_{i} + \pi_{0} = w(N), \\
		\sum_{i \in S} \pi_{i} + \pi_{0} \geq w(S) - \epsilon, S \in 2^N \backslash \emptyset \}.
	\end{aligned}
\end{equation}

In the concept of strong $\epsilon$-core \cite{driessen2013cooperative}, $\epsilon$ represents a minor perturbation or a threshold for deviation. If a coalition $S \subset N$ can increase its utility by more than the threshold, i.e., $\epsilon$, through deviating from $N$, then the coalition will choose to deviate and not collaborate with the participants in $N \backslash S$. The strong $\epsilon$-core of a game can be interpreted as the set of all allocation vectors that cannot be improved upon by any coalition if one imposes a cost of $\epsilon$ in all cases where a nontrivial coalition is formed \cite{driessen2013cooperative}. 

For convenience, the surplus vector in the strong $\epsilon$-core, $\pi$ is called core surplus vector, and the surplus vector determined by the VCG-like payment, $\pi^{VCG}$ is also called VCG surplus vector. The strong $\epsilon$-core is a set solution, while the VCG surplus vector is a single point solution. Obviously, increasing $\epsilon$ can expand the strong $\epsilon$-core. A sufficiently large $\epsilon$ can ensure the VCG surplus vector is in the strong $\epsilon$-core.

\begin{theorem}\label{thm2}
	Denote $\alpha(S) = \max_{T \supseteq S} \max_{i \in S} \Big[\big( w(N) - w(N\backslash \{i\}) \big) - \big( w(T) - w(T\backslash \{i\}) \big)\Big]$ for any coalition $S \subseteq N$. Given $\epsilon \geq 0$ satisfying $\epsilon \geq \max_{S} \alpha(S) (|N| - |S|)$, then the VCG surplus vector is in the strong $\epsilon$-core.
\end{theorem}
\begin{IEEEproof}
	The VCG surplus vector is defined as
	\begin{equation}
		\pi_i^{VCG} = w(N) - w(N\backslash \{i\}), \ \ \ \ i \in N.
	\end{equation}
	
	For any coalition $S = \{1, 2, \cdots, q\} \subseteq N$, we show that the blocking inequality associated with coalition $S$ is satisfied.
	\begin{equation}
		\begin{aligned}
			& \sum_{i \in S} \pi_{i}^{VCG} + \pi_{0}^{VCG}\\
			=& w(N) - \sum_{i = q+1}^{n} \pi_{i}^{VCG} \\
			=& w(N) - \sum_{i = q+1}^{n} \big( w(N) - w(N \backslash \{i\}) \big) \\
			\geq& w(N) - \sum_{i=q+1}^{n} \big( w(\{1,2\cdots,i\}) - w(\{1,2\cdots,i-1\}) + \\ 
			&\frac{\epsilon}{n - q}\big)\\
			=& w(N) - \big( w(\{1,2\cdots,n \}) - w(\{1,2\cdots,q \}) + \epsilon\big) \\
			=& w(S) - \epsilon.
		\end{aligned}
	\end{equation}
	
	Hence, the VCG surplus vector is in the strong $\epsilon$-core.
\end{IEEEproof}

The above theorem provides a lower bound on $\epsilon$ to ensure that the VCG surplus vector is in the strong $\epsilon$ core. If a very small lower bound can guarantee that the VCG surplus is in the core, we can directly employ the VCG-like payment to incentivize players. In the theorem mentioned above, this lower bound is difficult to calculate. Although we are uncertain whether the VCG surplus is in the $\epsilon$-core, we still have another way to comprehensively consider the VCG surplus and the strong $\epsilon$-core. 

\begin{lemma}\label{lemma1}
	For any player $i \in N$, the core surplus is not greater than $\pi_i^{VCG} + \epsilon$.
\end{lemma}
\begin{IEEEproof}
	Given any type profile $\hat{D}$, for any player $i$, $\pi_i \geq \pi_i^{VCG} + \epsilon$ holds, otherwise, there is at least one player $k \in N$, such that
	\begin{equation}
		\pi_k > \pi_k^{VCG} + \epsilon = w(N) - w(N \backslash \{k\}) + \epsilon,
	\end{equation}
	and we have the following inequality
	\begin{equation}
		\sum_{i \in N\backslash\{k\}} \pi_i + \pi_0 = w(N) - \pi_{k} < w(N \backslash\{k\}) - \epsilon,
	\end{equation}
	which contradicts the core constraint.
\end{IEEEproof}

\begin{theorem}\label{thm4}
	For the core payment $p_i = \pi_i - v_i(F(\hat{D}),\hat{D}_i)$, the amount that player $i$ can benefit by deviating from the truthful input strategy is less than or equal to $p_i^{VCG}(F(D)) - p_i(F(D)) + \epsilon$.
\end{theorem}
\begin{IEEEproof}
	Suppose not, there is some input $\hat{D}_i$ such that
	\begin{equation}
		\begin{aligned}
			&\Big( v_i(F(\hat{D}_i \cup D_{-i}),D_i) + p_i(F(\hat{D}_i \cup D_{-i})) \Big) - \\ &\Big( v_i(F(D),D_i) +  p_i(F(D)) \Big) \\ &> p_i^{VCG}(F(D)) - p_i(F(D)) + \epsilon,
		\end{aligned}
	\end{equation}
	where $p_i(F(D))$ is the payment when the player inputs truthfully.
	
	Next, we have
	\begin{equation}
		\begin{aligned}
			&v_i(F(\hat{D}_{i} \cup D_{-i}),D_i) + p_i(F(\hat{D}_{i} \cup D_{-i})) \\ &> v_i(F(D),D_i) + p_i^{VCG}(F(D)) + \epsilon.
		\end{aligned}
	\end{equation}
	
	Note that for any input, we have $v_i(F(\hat{D}_{i} \cup D_{-i}),D_i) + p_i(F(\hat{D}_{i} \cup D_{-i})) \leq v_i(F(\hat{D}_{i} \cup D_{-i}),D_i) + p_i^{VCG}(F(\hat{D}_{i} \cup D_{-i})) + \epsilon$ for all $i \in N$ due to Lemma \ref{lemma1}. So we can enlarge the left part of the inequality as follows,
	\begin{equation}
		\begin{aligned}
			&v_i(F(\hat{D}_{i} \cup D_{-i}),D_i) + p_i^{VCG}(F(\hat{D}_{i} \cup D_{-i})) + \epsilon> \\ &v_i(F(D),D_i) + 
			p_i^{VCG}(F(D)) + \epsilon.
		\end{aligned}
	\end{equation}
	
	The left part of the above inequality is the utility when the player inputs untruthfully, and the right part is the utility when the player inputs truthfully. The inequality contradicts the incentive compatibility property of the VCG-like payments.
\end{IEEEproof}

Theorem \ref{thm4} shows that by reducing the distance between $p_i$ and $p_i^{VCG} + \epsilon$, we can limit the benefits of inputting false data. In other words, although we cannot find a VCG surplus vector in the strong $\epsilon$-core, reducing the distance between $p_{i}$ and $p_{i}^{VCG} + \epsilon$ as much as possible can still motivate participants to input data truthfully. Eventually, our core-selecting incentive mechanism for federated learning is proposed as follows. 

\begin{enumerate}[Step 1:]
	\item Before computing the payments, model $F(\hat{D}_S)$ is trained and evaluated for each coalition $S \subseteq N$. Meanwhile, the characteristic values are calculated as 
	\begin{equation}
		w(S) = \sum_{i\in S}v_{i}(F(\hat{D}_S),\hat{D}_i) + b_0, \ \ \ \ S \subseteq N.
	\end{equation}
	\item Next, the VCG surplus is determined as
	\begin{equation}
		\pi_i^{VCG} = w(N) - w(N\backslash \{i\}), \ \ \ \ i \in N.
	\end{equation}
	
	\item The following quadratic programming is solved to give a solution $\pi^{*}$.
	\begin{equation}\label{eps core-selecting}
		\begin{aligned}
			\min & \ \ \sigma^2 = \sum_{i \in N} (\pi_i - \pi_i^{VCG} + \epsilon)^2 \\
			\text{s.t.} & \ \ \epsilon \geq 0, \\
			& \ \ \sum_{i \in N} \pi_i + \pi_0 = w(N), \\
			& \ \ \sum_{i \in S} \pi_i + \pi_0 + \epsilon \geq w(S), \ \ \ \ &S \subseteq N,\\
			& \ \ \pi_i \geq 0, &i \in N \cup \{0\},\\
		\end{aligned}
	\end{equation}
	\item Finally, the payments given to participants are calculated as
	\begin{equation}
		p_i = \pi_i^{*} - v_i(F(\hat{D}), \hat{D}_i), \ \ \ \ i \in N.
	\end{equation}
\end{enumerate}

The above procedure determines the observable surplus vector that we want to achieve. In particular, if the VCG surplus vector is in the $\epsilon$-core, the solution of the core-selecting incentive mechanism is equal to the VCG surplus vector. Conservative participants are prefer to input data truthfully because the payment under the core-selecting mechanism may be equal to or very close to that under the truthful incentive mechanism. Even if aggressive participants attempt to input false data, Theorem \ref{thm4} indicates that the additional benefits obtained by inputting false data are also minimized. Since the selected surplus vector is taken from the strong $\epsilon$-core and $\epsilon$ is also reduced in the programming, there is little incentive for any coalition or any participant to deviate from cooperation.

\section{Efficient Core-selecting Incentive Mechanism based on Sampling Approximation}\label{Efficient Core-selecting Incentive Mechanism based on Sampling Approximation and Reputation}
The proposed core-selecting mechanism is an ideal mechanism for federated learning. However, exact computation of the core-selecting mechanism requires evaluating the additional model with $2^n$ times to obtain the characteristic values before solving the quadratic programming, which is computationally expensive. To address this, our method is to sample a relatively small number of coalitions from a probability distribution, and compute the desired solution on the sampled coalitions. 

In our method, a set of coalition $\mathcal{S} =\{S_1, S_2, \cdots, S_m\}$ is sampled from a probability distribution $\mathcal{D}$ and the characteristic values of these coalitions are given by aggregating and evaluating $m$ additional models. Then, solving the core-selecting quadratic programming over $m$ coalitions instead of $2^n$ coalitions gives an approximate core surplus vector, resulting in the following quadratic programming.
\begin{equation}\label{approximate quadratic programming}
    \begin{aligned}
    \min & \ \ \sigma^2 = \sum_{i \in N} (\pi_i - \pi_i^{VCG} + \epsilon)^2 \\
    \text{s.t.} & \ \ \sum_{i \in N} \pi_i + \pi_0 = w(N), \\
    & \ \ \sum_{i \in S} \pi_i + \pi_0 + \epsilon \geq w(S_k), \ \ \ \ & k=1,2,\cdots,m,\\
    & \ \ \pi_i \geq 0, &i \in N \cup \{0\}.\\
    \end{aligned}
\end{equation}

The solution $\hat{\pi}$ of the above quadratic programming may not satisfy all $2^n$ core constraints. In order to analyze its properties, we consider that $\hat{\pi}$ satisfies each core constraint with a probability, and similar to \cite{yan2021if}, we introduce the definition of $\delta$-probable core as follows.
\begin{myDef}[$\delta$-probable core]
	A surplus vector $\pi$ is in the $\delta$-probable core if and only if
	\begin{equation}
	\mathop{\text{Pr}}\limits_{S \sim \mathcal{D}} \Big[\sum_{i \in S}\pi_i + \pi_0  + \epsilon - v(S) \geq 0\Big] \geq 1 - \delta.
	\end{equation}
\end{myDef}

Given any coalition $S$ drawn from $\mathcal{D}$, the core constraint is violated with probability at most $\delta$. For the exact core-selecting surplus $\pi^*$, we have
\begin{equation}
    \mathop{\text{Pr}}\limits_{S \sim \mathcal{D}} \Big[\sum_{i \in S}\pi_i^* + \pi_0^* + \epsilon^{*} - v(S) \geq 0\Big] = 1.
\end{equation}

The definition of $\delta$-probable core can explain the constraints of approximate quadratic programming (\ref{approximate quadratic programming}). The next question we focus on is how to determine the sample size $m$. The following theorem \ref{thm5} answers this question. Before presenting this theorem, we introduce two known lemmas \cite{Shalev2014Understanding}.

\begin{lemma}\label{lemma2}
    Let $\mathcal{F}$ be a function class from $\mathcal{X}$ to $\{-1, 1\}$, and let $y$ be the true value function. If $\mathcal{G}$ has VC-dimension $d$, then with
    \begin{equation}
    m = O \Big( \frac{d + \log( \frac{1}{\Delta}) }{\delta^2}\Big),
    \end{equation}
    i.i.d. samples $\text{x}^1, \cdots, \text{x}^m \sim \mathcal{P}$, we have
    \begin{equation}
    \Big| \mathop{\text{Pr}}\limits_{\text{x}\sim \mathcal{P}} [f(\text{x}) \neq y(\text{x})] - \frac{1}{m}\sum_{i = 1}^{m} \mathds{1}_{f(\text{x}^i) \neq y(\text{x}^i)}\Big| \leq \delta,
    \end{equation} 
    for all $f \in \mathcal{F}$ and with probability $1 - \Delta$.
\end{lemma}

\begin{lemma}\label{lemma3}
    The function class $\mathcal{F}^{n} = \{\text{x} \mapsto \text{sign}(\text{w} \cdot \text{x}): \text{w} \in \mathbb{R}^n \}$ has VC-dimension $n$.
\end{lemma}
%working

\begin{theorem}\label{thm5}
    Given a distribution $\mathcal{P}$ over $2^N$, and $\delta$, $\Delta>0$, solving the programming (\ref{approximate quadratic programming}) over $O((n+\log(1/\Delta))/\delta^2)$ coalitions sampled from $\mathcal{P}$ gives a surplus vector in the $\delta$-probable core with probability $1 - \Delta$.
\end{theorem}
\begin{IEEEproof}
    Given a coalition $S$ sampled from $\mathcal{P}$, we convert it into a vector $\text{z}^S = (z^S, -w(S), 1, 1)$ where $z \in \{0,1\}^n$ and $x^S_i = 1$ if $i \in S$ and $x^S_i = 0$ otherwise. 

    Consider a linear classifier $f$ defined by parameter $\text{w}^f = (\pi, 1, \pi_0, \epsilon)$ where $\text{w}^f \in \mathbb{R}^{n+3}$. If $\text{sign}(\text{w}^f \cdot \text{z}^{S}) = 1$, then we have $\text{w}^f \cdot \text{z}^{S} = \sum_{i \in S}\pi_{i} + \pi_0 + \epsilon - w(S) \geq 0$. Obviously, classifier $f(\text{z}^S) = \text{sign}(\text{w}^f \cdot \text{z}^{S})$ can identify the core constraint for coalition $S \subseteq N$. If a linear classifier $f$ satisfies $f(\text{z}^S) = 1$ for all coalitions $S \subseteq N$, then it represents a surplus vector in the $\epsilon$-core . This inspires us to define a class of functions to represent the $\delta$-probable core:
    \begin{equation}
    \begin{aligned}
            \mathcal{F} = \Big\{ \text{z} \mapsto \text{sign}(\text{w} \cdot \text{z}):\text{w} = (\pi,1,\pi_0,\epsilon), \pi \in \mathbb{R}^n_+, \\ 
    \pi_0 \geq 0, \sum_{i = 1}^{n}\pi_i +\pi_0 = w(N) \Big\}.
    \end{aligned}
    \end{equation}

    This class of functions $\mathcal{F}$ is a subset of $\mathcal{F}^{n+3}$, and it has VC-dimension at most $n+3$ by Lemma \ref{lemma3}.

    Suppose that we solve quadratic programming (\ref{approximate quadratic programming}) on coalition samples $S_1, \cdots, S_m$, which gives us a solution $(\hat{\pi}, \hat{\pi}_0, \hat{\epsilon})$ and the corresponding classifier $\hat{f}$. Note that $\hat{f}(\text{z}^{S_k}) = 1$ holds for $k = 1, \cdots, m$.

    By Lemma \ref{lemma2}, the following inequality holds with probability $1 - \Delta$.
    \begin{equation}
        \begin{aligned}
        &\mathop{\text{Pr}}\limits_{S \sim \mathcal{P}} \Big[\sum_{i \in S}\hat{\pi}_i + \hat{\pi}_0 + \hat{\epsilon} - w(S)  \geq 0\Big] \\
        &= 1 - \mathop{\text{Pr}}\limits_{S \sim \mathcal{P}} [\hat{f}(\text{z}^S) \neq y(\text{z}^S)] \\
        &= 1 - \Big( \mathop{\text{Pr}}\limits_{S \sim \mathcal{P}} [\hat{f}(\text{z}^S) \neq y(\text{z}^S)] - \frac{1}{m}\sum_{k = 1}^{m} \mathds{1}_{\hat{f}(\text{x}^{S_k}) \neq y(\text{x}^{S_k})}\Big) \\
        &\geq 1 - \delta,
        \end{aligned}
    \end{equation}
    where $\mathcal{P}$ is a $2^n$ dimensional probability distribution and the second transition holds because $\hat{f}$ and $y$ agree on $S_1, \cdots, S_m$.
\end{IEEEproof}

Theorem \ref{thm5} provides a theoretical result for our approximate method. The above incentives reasonably evaluate the contributions of participants. The VCG surplus can be seen as a marginal contribution, and surplus $\hat{\pi}$ can be seen as a contribution based on the approximate strong $\epsilon$-core. 

Next, in order to aggregate and obtain a better model, we employ a non-uniform weighted aggregation scheme. Reputation is widely used to measure the reliability of a participant based on its past behavior \cite{kang2018blockchain, huang2018software, shi2022fedfaim}. The participant with a better reputation will be assigned a greater weight. At each round $t$, the surplus $\hat{\pi}_i^t$ can be determined. We calculate the reputation of each participant $i$ based on its contribution from round $1$ to round $t$. Let $R_i^t$ denote participant $i$’s reputation. It can be computed as
\begin{equation}
	R_i = \max (\phi_0, \sum_{\tau=1}^{t}\hat{\pi}_i^t),
\end{equation}
where $\phi_0$ is a small positive number representing the lower bound. In the next round $t+1$, we will use $R_i^t$ to aggregate the new global model as
\begin{equation}
	F^{t+1}(D) = \sum_{i \in N} \frac{R_i^t}{\sum_{j \in N}R_j^t} F^{t+1}(D_i).
\end{equation}

We implement the efficient core-selecting incentive mechanism by adding additional model evaluation, and the computation process is shown in Algorithm \ref{FedAvgIncentive}. 

\begin{algorithm}[htbp]
	\caption{FedAvg Core-selecting Incentive Algorithm}
	\begin{algorithmic}
		\State \textbf{Input:} Participants $N$ and their data $\{D_{i}\}_{i \in N}$, number of coalition samples $m$.
		\State \textbf{Output:} Accumulated payment $P_i$.\\
		
		\State Initialize global model $\theta^0$, reputation $R_i \gets \phi_0$ and accumulated payment $P_i \gets 0$.
		\For{each round $t = 1,2,\cdots,T$}
		\For{each participant $i \in N$ \textbf{in parallel}}
		\State $\theta_{i}^{t} = \text{ParticipantUpdate}(i, \theta^{t-1})$
		\EndFor
		\State // \emph{Aggregate and Evaluate}
		\State $\mathcal{S} \leftarrow$ randomly select $m$ coalitions from $2^{N} \backslash \emptyset$
		\For{each coalition $S \in \mathcal{S} \cup \{N\} \cup \{N \backslash \{i\} \}_{i \in N} $}
		\State Aggregate $\theta_{S}^{t} \gets \sum_{i \in S} \frac{R_i}{\sum_{j \in S} R_j} \theta_{i}^{t}$
		\State Evaluate the accuracy of model $\theta_{S}^{t}$ and compute characteristic value $w(S)$ as (\ref{characteristic function})
		\EndFor
		\State Update global model $\theta^t = \theta_{N}^{t}$
		\State Compute $\pi_i^{VCG} = w(N) - w(N\backslash i)$ for $i \in N$ and solve the programming as (\ref{approximate quadratic programming}) to give a solution $\hat{\pi}$.
		\State Each participant $i \in N$ receive $p_i = \hat{\pi} - v_i$.
		\State Update $P_i \gets P_i + p_i$ and $R_i \gets R_i + \hat{\pi}_i$
		
		\EndFor
		\State \textbf{$\text{ParticipantUpdate}(i, \theta)$} \ \ \ \ // \emph{Run on participant $i$}
		\State $Z \leftarrow$ (split $D_i$ into batches of size $B$)
		\For{each local epoch $e = 1, \cdots, E$}
		\For{batch $z \in Z$}
		\State $\theta \leftarrow \theta - \eta \nabla \mathcal{L}(\theta; z)$
		\EndFor
		\EndFor
		\State return $w$ to server
	\end{algorithmic}
	\label{FedAvgIncentive}
\end{algorithm}

\section{Experiments}\label{Experiments}%checked

The purpose of this section is threefold. First, we empirically demonstrate our theoretical results. In the second part of the experiment, we investigate the influence of inputting untruthful data on the incentive mechanism based on VCG-like payments, core-selecting payments, and efficient core-selecting payments. Moreover, we will test the efficiency of our mechanism.

All tests are performed on a computer with an Intel Xeon 5218 CPU, Nvidia Quadro RTX4000 GPU and 64GB RAM. The optimization problems are solved with Gurobi.

\subsection{Validation} %checked
\begin{figure}[htbp]
	\centering
	\includegraphics[width=2.5in]{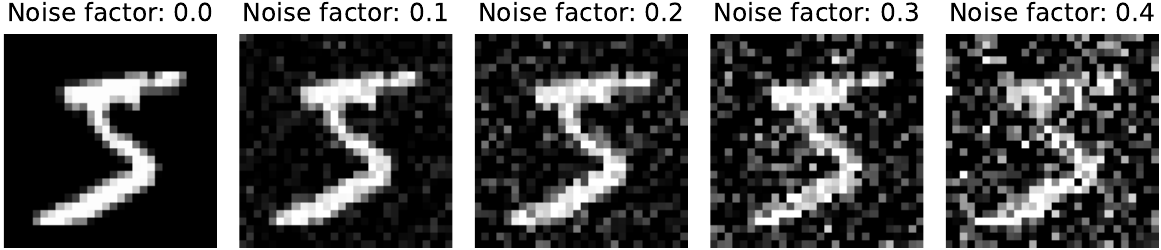}
	\caption{Data with noise.}
	\label{diag_adding_noise}
\end{figure}

\begin{figure}[htbp]
	\centering
	\includegraphics[width=2.5in]{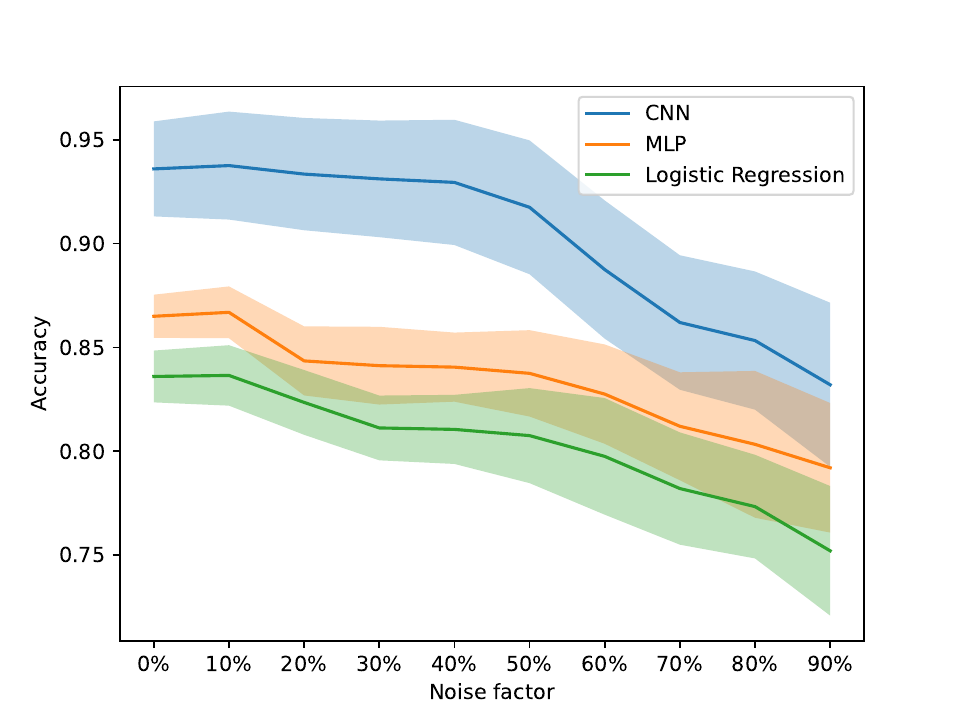}
	\caption{Accuracy of inputting noisy data.}
	\label{Accuracy of inputting noisy data}
\end{figure}

To verify our assumption, a participant is selected, and the participant will input noisy data as shown in Fig. \ref{diag_adding_noise}. The MNIST dataset is chosen. 10 percent of the dataset is set as an independent test set for the server, and the remaining is divided into 5 parts for 5 participants. Based on Algorithm \ref{FedAvg}, the logistic regression, multi-layer perceptron (MLP), and convolutional neural network (CNN) models are trained for 10 rounds, respectively. The results are shown in Fig. \ref{Accuracy of inputting noisy data}. We can see that the accuracy mainly decreases as the false degree increases. Due to noise causing overfitting, the accuracy increases by around 10 percent. For this phenomenon, we believe that this type of data is also high-quality, as long as it can improve the performance of the global model.

To verify theorem \ref{thm5}, we randomly sample a small fraction of coalitions to implement the efficient core-selecting mechanism for a single round and determine what fraction of all coalitions satisfy the core constraints, which gives us the fraction $1-\delta$, referred to as core accuracy. In this experiment, we chose the smaller-scale iris dataset that only has 4 features. This makes it computationally feasible to evaluate all possible models and compute the core-selecting payment exactly. 10 percent of the dataset is set up as an independent test set for the server, and the remaining is divided into 10 parts for 10 participants. Based on Algorithm \ref{FedAvgIncentive}, we will train the logistic regression models and multi-layer perceptron (MLP) models using stochastic gradient descent. The parameters are set as $b_0 = 2$ and $k_i = 2$. The payment, utility, and error will be calculated. 

\begin{figure}[htbp]
\centering
\includegraphics[width=2.5in]{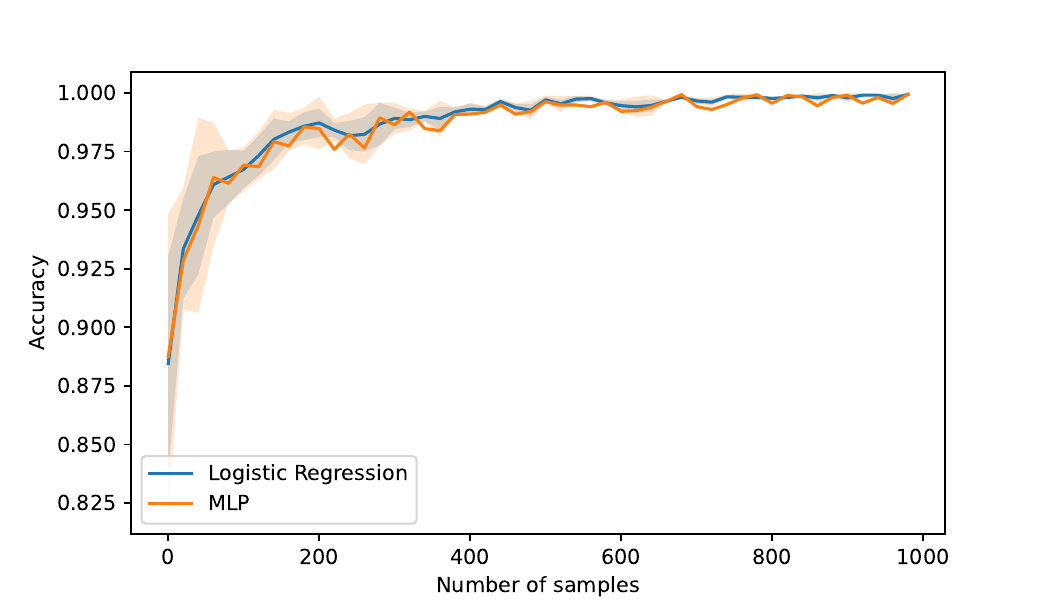}
\caption{Core accuracy with different number of samples.}
\label{fig_A1}
\end{figure}

As shown in Fig. \ref{fig_A1}, even with a small number of sampled coalitions, the resultant surplus vector is still in the $\delta$-probable core. 

\begin{figure}[htbp]
	\centering
	\includegraphics[width=2.5in]{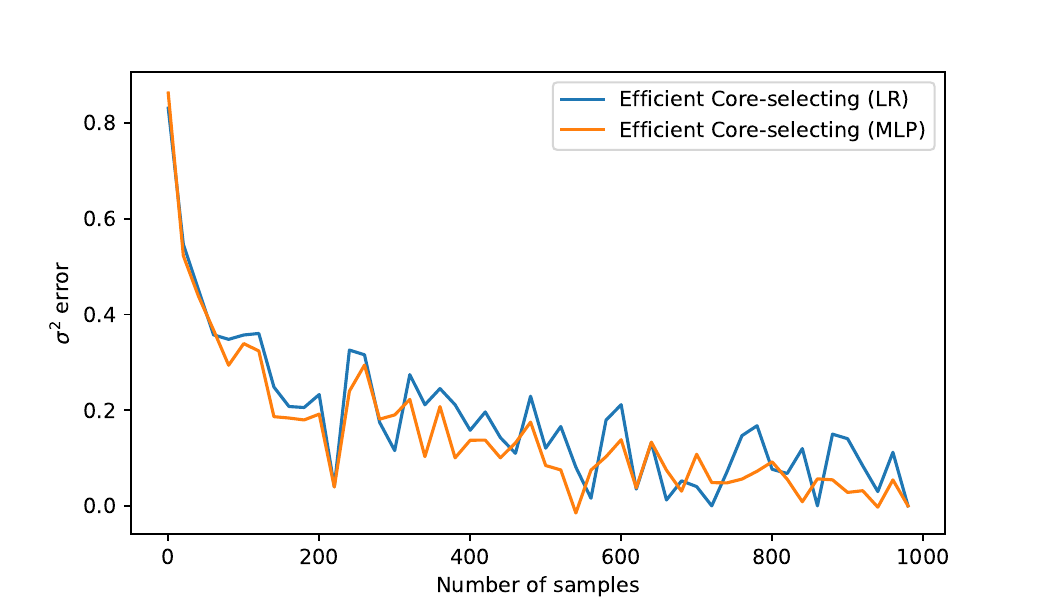}
	\caption{$\sigma^2$ value with different number of samples.}
	\label{fig_A2}
\end{figure}

Then, as shown in Fig. \ref{fig_A2}, the errors of $\sigma ^ 2$ decrease as the number of samples increases. 

\begin{figure}[htbp]
	\centering
	\includegraphics[width=2.5in]{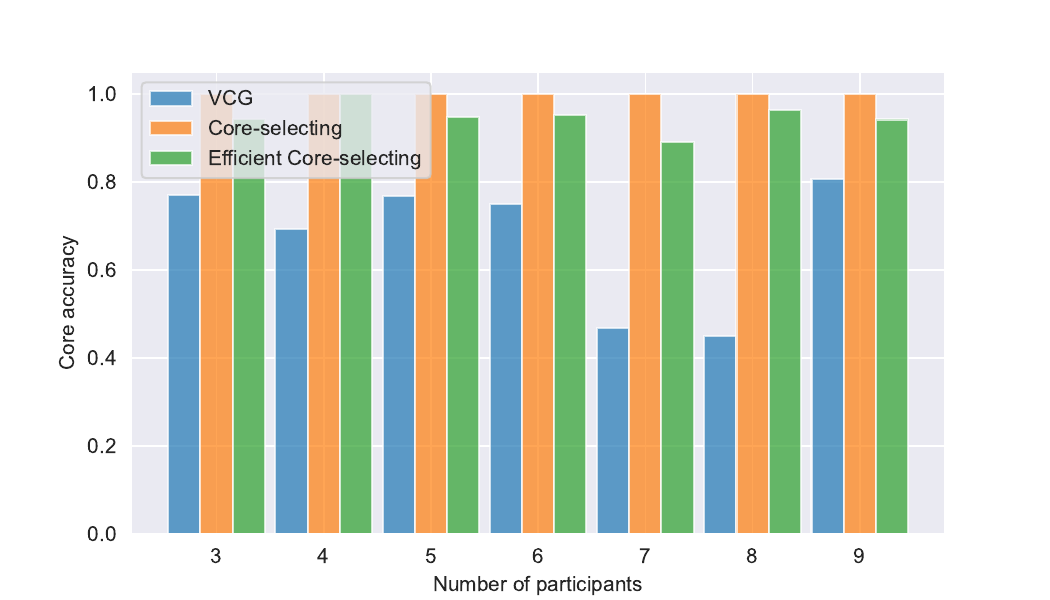}
	\caption{Core accuracy with different number of participants (Logistic Regression, $\delta = 0.3, \Delta = 0.3$).}
	\label{fig_D1}
\end{figure}

Next, as shown in Fig. \ref{fig_D1}, we can see that the VCG-like payment cannot satisfy all core constraints, and the efficient core-selecting payment has a higher average accuracy compared to the VCG-like payment. 

The above experiments show that the difference between the efficient core-selecting payment and the exact core-selecting payment decreases as the number of samples increases. When the scale of the problem is large, the efficient core-selecting mechanism can improve the computational efficiency by reducing the number of samples. 

\subsection{Truthfulness}
In this experiment, we compare the effects of different inputs on the incentive mechanism based on the VCG-like, core-selecting, and efficient core-selecting payments. Of course, the three payments rule also corresponds to three types of aggregation weights. We still use the MNIST dataset and divide it into 5 parts for 5 participants. The parameter settings from the previous experiment. The logistic regression, multi-layer perceptron (MLP), and convolutional neural network (CNN) models are trained for 10 rounds, respectively. The parameters of the efficient core-selecting are set as $\delta = 0.5$ and $\Delta = 0.5$. To simulate the input behavior of participants, we select a participant and design three strategies for the participant to input data untruthfully. Then the selected participant's accumulated utility is computed. We repeat the entire process 10 times to compute the mean utility.

\begin{figure}[htbp]
	\centering
	\includegraphics[width=2.5in]{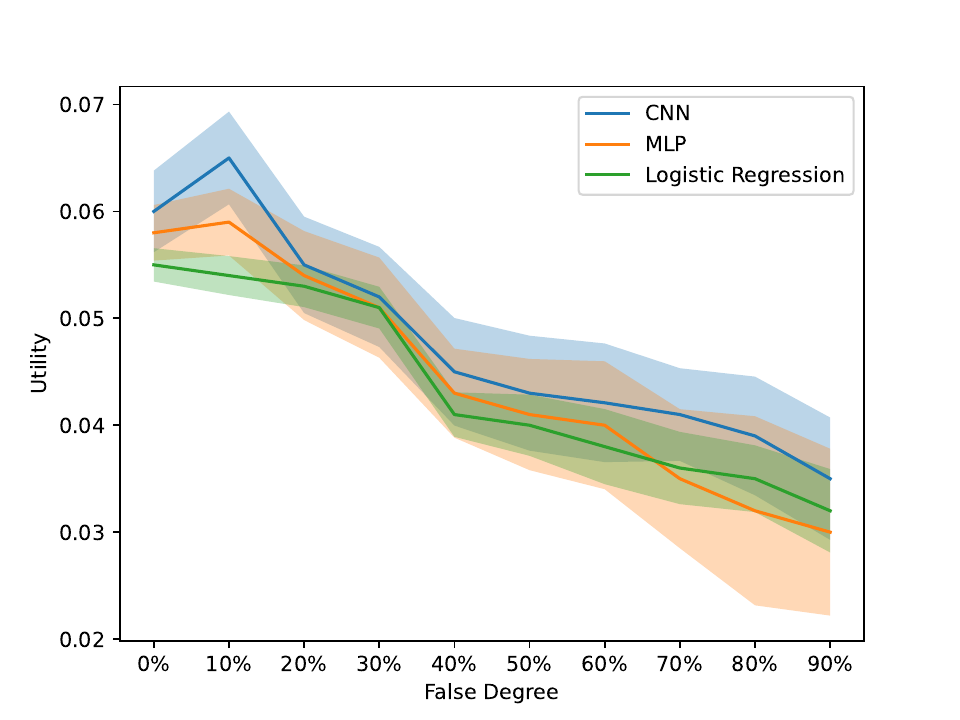}
	\caption{Actual utility (adding noise).}
	\label{fig_B1}
\end{figure}

The first input strategy is adding different proportion of white noise to all training data. The utility of the selected participant is shown in Fig. \ref{fig_B1}. It is easy to see that the utility decreases with the increase in false degrees. In particular, when the noise is increased to 10 percent, the utility does not decrease but increases slightly. The reason is that adding noise to the data during the training process can prevent overfitting, which is commonly used as a regularization method.

\begin{figure}[htbp]
	\centering
	\subfloat[data removal.]{\includegraphics[width=1.7in]{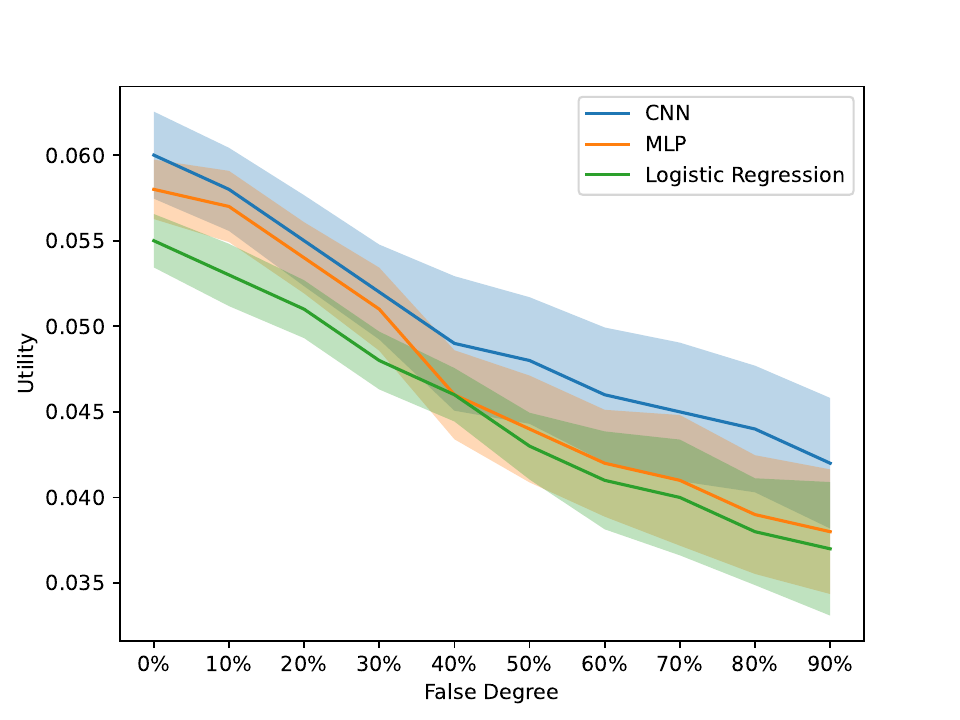}%
			}
	\hfil
	\subfloat[incorrect label.]{\includegraphics[width=1.7in]{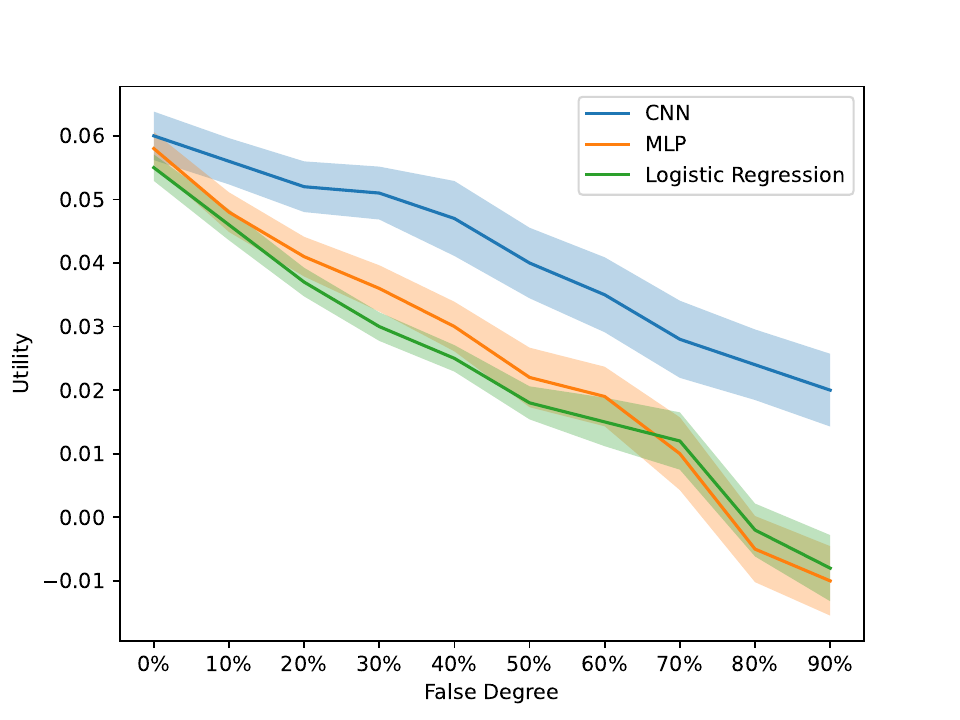}%
			}
	\caption{Actual utility of efficient core-selecting incentive mechanism.}
	\label{fig:two_images}
\end{figure}

The second input strategy is removing a proportion of data, which is similar to the work of \cite{yan2021if}. The third input strategy is inputting some incorrect labels, which is malicious behavior. The results are shown in Fig. \ref{fig:two_images}, and we can see that the utility of the third strategy decreases more significantly than other strategies because it is more malicious than other strategies.

Because rational participants will maximize the utility, the above experiments show that the efficient core-selecting mechanism can prevent participants from inputting low-quality or false data. 

\begin{figure}[htbp]
	\centering
	\subfloat[data removal.]{\includegraphics[width=1.7in]{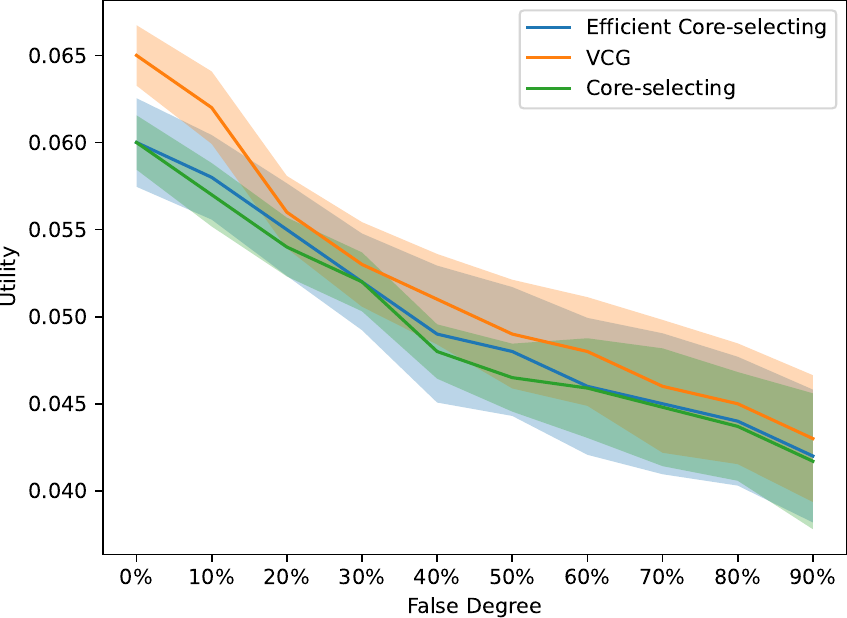}%
		}
	\hfil
	\subfloat[incorrect label.]{\includegraphics[width=1.7in]{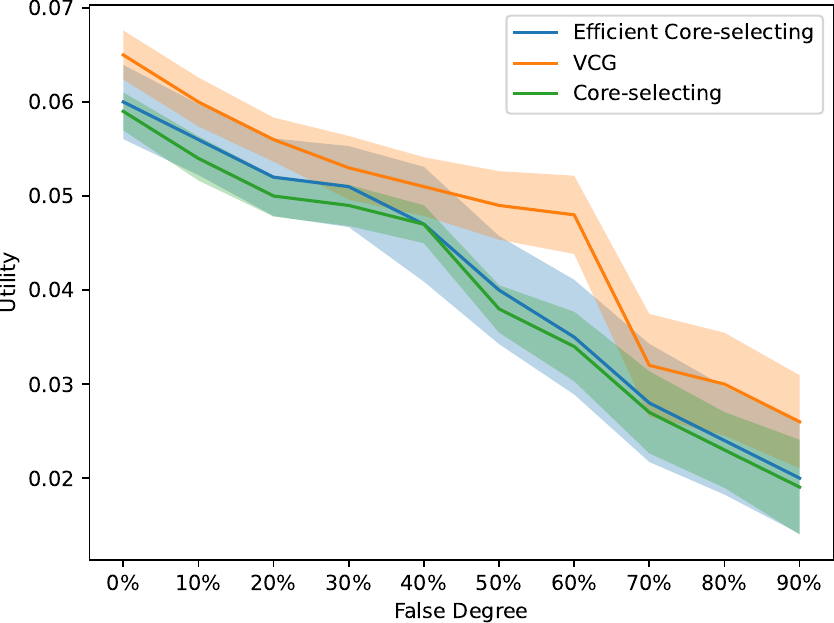}%
		}
	\caption{Actual utility of three mechanisms (CNN, $\delta = 0.5$, $\Delta = 0.5$).}
	\label{fig:two_images_different mechanisms}
\end{figure}

Additional experiments are conducted to compare the three mechanisms. As shown in Fig. \ref{fig:two_images_different mechanisms}, we can see that there is not much difference between the core-selecting mechanism and the efficient core-selecting mechanism. 

\subsection{Efficiency}

\begin{figure}[htbp]
	\centering
	\includegraphics[width=2.5in]{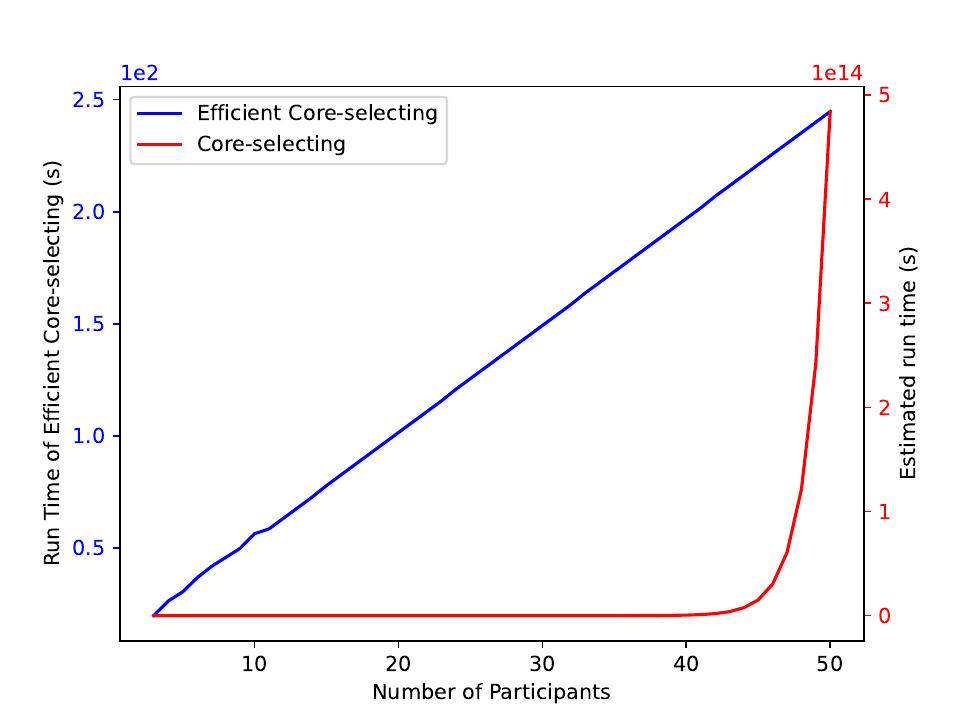}
	\caption{Run time with different number of participant (We only calculated the run time for $n \leq 10$, while the run time for $n>10$ is estimated).}
	\label{fig_C3}
\end{figure}

In this experiment, we visualize the run time as the number of people increases. We still use the MNIST dataset. The parameters of the efficient core-selecting mechanism are fixed to sample coalitions ($\delta=0.3$ and $\Delta=0.3$). The main computational overhead is that additional models need to be aggregated and evaluated in each round, so we only compare the running time of a single round. As shown in Fig. \ref{fig_C3}, as the number of participants increases, the run time of the efficient core-selecting mechanism increases linearly, while the run time of the core-selecting mechanism increases exponentially. 

Therefore, the efficient core-selecting mechanism can reduce computational overhead compared to the core-selecting mechanism. 

\section{Conclusion}\label{Conclusion}% checked
In this paper, in order to motivate participants to input truthful data and promote stable cooperation in federated learning, we propose an efficient core-selecting incentive mechanism. First, we introduce a data sharing game for federated learning. Then we used a core-selecting incentive mechanism that combines the advantages of both the VCG-like payment and the core. Different from the classical core-selecting mechanism, this mechanism adopts a relaxation on the core to deal with the randomness in federated learning and minimizes the benefits of inputting false data, which can promote stable cooperation among players and also evaluate the quality of participants' data. Since the core-selecting incentive mechanism requires exponential time to aggregate and evaluate additional models, we propose an efficient core-selecting mechanism based on sampling approximation. To avoid the impact of low-quality data, the mechanism adjusts the aggregation weight of participants based on their historical contributions. Our experiments demonstrate that the proposed mechanism can prevent participants from inputting false data. In addition, participants will not deviate from cooperation because this mechanism can achieve the desired core accuracy through sampling. 

For future work, we wish to continue exploring the connection between federated learning and game theory, and develop more reasonable payments for participants. Additionally, combining reputation and participant selection may improve the performance of global model.

\bibliographystyle{IEEEtran}
\bibliography{reference}

\end{document}